\documentclass[apj]{emulateapj}
\usepackage{graphicx}
\usepackage{amsmath,amssymb}
\usepackage{amsmath}

\let\oldAA\AA
\renewcommand{\AA}{\text{\normalfont\oldAA}}

\begin{document}

\title{The Supernova Rate Beyond The Optical Radius}

\author{Sukanya Chakrabarti,\altaffilmark{1} 
Brennan Dell,\altaffilmark{1}
Or Graur,\altaffilmark{2,3,4} 
Alexei V. Filippenko,\altaffilmark{5,6}
Benjamin T. Lewis, \altaffilmark{1}
and Christopher F. McKee \altaffilmark{7}}

\altaffiltext{1}
{School of Physics and Astronomy, Rochester Institute of Technology, 84 Lomb Memorial Drive, Rochester, NY 14623; chakrabarti@astro.rit.edu}
\altaffiltext{2}
{Harvard-Smithsonian Center for Astrophysics, 60 Garden St., Cambridge, MA 02138, USA }
\altaffiltext{3}
{Department of Astrophysics, American Museum of Natural History, New York, NY 10024, USA}
\altaffiltext{4}
{NSF Astronomy and Astrophysics Postdoctoral Fellow}
\altaffiltext{5}
{Department of Astronomy, 501 Campbell Hall, University of California, Berkeley, CA 94720-3411}
\altaffiltext{6}
{Miller Senior Fellow, Miller Institute for Basic Research in Science, University of California, Berkeley, CA 94720}
\altaffiltext{7}
{Departments of Physics and of Astronomy, University of California, Berkeley CA 94720}

\begin{abstract}
Many spiral galaxies have extended outer H~I disks and display low levels of star formation, inferred from the far-ultraviolet emission detected by {\it GALEX}, well beyond the optical radius. Here, we investigate the supernova (SN) rate in the outskirts of galaxies, using the largest and most homogeneous set of nearby supernovae (SNe) from the Lick Observatory Supernova Search (LOSS).  While SN rates have been measured with respect to various galaxy properties, such as stellar mass and metallicity, their relative frequency in the outskirts versus the inner regions has not yet been studied.  Understanding the SN rate as a function of intragalactic environment has many ramifications, including the interpretation of LIGO observations, the formation of massive stars, and the puzzlingly high velocity dispersion of the outer H~I disk.  Using data from the LOSS survey, we find that the rate beyond the optical radius of spiral galaxies is $2.5 \pm 0.5$ SNe per millennium, while dwarf galaxies host $ 4.0 \pm 2.2$ SNe per millennium.  The rates of core-collapse SNe (that may collapse to form the massive black holes detected by LIGO/Virgo) in the outer disks of spirals is $1.5 \pm 0.15$ SNe per millennium and in dwarf galaxies is $2.6 \pm 1.5$ SNe per millennium.  Core-collapse SNe in spiral outskirts contribute $7600 \pm 1700$\,SNe\,~Gpc$^{-3}$\,yr$^{-1}$ to the volumetric rate, and dwarf galaxies have a rate of $31,000 \pm 18,000$\,SNe\,Gpc$^{-3}$\,yr$^{-1}$.  The relative ratio of core-collapse to Type Ia SNe is comparable in the inner and outer parts of spirals, and in dwarf galaxies.  
\end{abstract}

\section{Introduction}

The rate of supernovae (SNe) is a fundamental quantity that has far-reaching implications for many areas of astrophysics and cosmology.  The supernova (SN) rate is directly tied to the metal enrichment of galaxies, as well as to the birth rate of neutron stars and black holes. The Lick Observatory Supernova Search (LOSS; Filippenko et al. 2001; Filippenko 2005; Leaman et al. 2011) with the 0.76-m Katzman Automatic Imaging Telescope (KAIT) has produced the largest homogeneous sample of relatively nearby SNe currently available for SN rate calculations. It has been analyzed to search for correlations between the SN rate and galaxy properties (Li et al. 2011b,c; Maoz et al. 2011; Graur et al. 2017a,b). Many spiral galaxies have extended H~I disks (Walter et al. 2008) and display far-ultraviolet emission in the outskirts (Thilker et al. 2007; Bigiel et al. 2010) that is indicative of a low level of star formation beyond the optical radius. Some of these stars will explode as SNe, but the SN rate in the outskirts of galaxies has not yet been determined.  Here, we analyze the LOSS sample to directly determine the SN rate beyond the optical radius.

LIGO observations (Abbott et al. 2016a,b,c; Abbott et al. 2017) indicate that merging massive ($\ga 10\,M_{\odot}$) binary black holes (BBHs) likely form in regions of low metallicity (Belczynski et al. 2016).  Recent work has studied the host galaxies of BBH mergers using binary merger population synthesis models (Lamberts et al. 2016; O'Shaughnessy et al. 2017; Chakrabarti et al. 2017).  Lamberts et al. (2016) showed that these events are either arising from dwarf galaxies or massive galaxies at high redshift.  Dwarf galaxies are observed to have low metallicities (Kirby et al. 2013; Lee et al. 2006).  O'Shaughnessy et al. (2017) noted that owing to the low-metallicity star formation found in dwarf galaxies, BBH mergers would be abundantly produced there.  

Chakrabarti et al. (2017) found that the outskirts of spiral galaxies, which also manifest a low level of star formation and have low metallicities (Kennicutt et al. 2003; Bresolin et al. 2009; Bigiel et al. 2010), may have a comparable contribution to the observed LIGO/Virgo detection rates as dwarf galaxies do.  If so, one may expect that not only black holes, but other tracers of dying massive stars, such as SNe, could be prevalent in the outer disks of spirals.  Future studies may be able to statistically identify the host galaxies of BBH mergers (Chen \& Holz 2016; Schutz 1986).  Especially for compact object mergers having electromagnetic counterparts, finding their host galaxies is more easily done for larger galaxies than for small, faint, dwarf galaxies.  

Of the types of SNe that have been observed by LOSS, we are particularly interested in core-collapse (CC) SNe, (Filippenko 1997; Matheson et al. 2001; Modjaz et al. 2014; Graur et al. 2017) which arise from the core collapse of stars more massive than $\sim 8\,M_{\odot}$.  To address the question of whether the LIGO/Virgo detections of massive black holes may in fact arise from spiral outskirts as often as they do in dwarf galaxies, we look at the rate of core-collapse SNe in the outskirts of spiral galaxies and contrast this with that found in dwarf galaxies. Type Ia SNe are understood to result from thermonuclear explosions of carbon-oxygen white dwarfs from stars having initial mass $< 8\,M_{\odot}$ (e.g., Maoz et al. 2014).  We compare the relative ratio of CC SNe to the lower-mass SNe~Ia in both spiral-galaxy outskirts and dwarf galaxies, and find that they are comparable to within the uncertainties.  

This paper is organized as follows.   In \S 2, we briefly review the parameters of the LOSS survey and summarize how we calculate SN rates.  We present our main results in \S 3, and we contrast in particular the SN rate in the outskirts of spiral galaxies with the rate in dwarf galaxies as a whole.   We also discuss the observed ratio of SNe~Ia to CC SNe in the outer disks of spiral galaxies and in dwarf galaxies, and the rates of CC SNe.  In \S 4, we discuss our results and conclude.

\section{Methods}

The LOSS sample is magnitude limited to $\sim 19$ mag, contains 14,878 galaxies, with 929 SNe (Leaman et al. 2011). Galaxies in the sample are grouped by Hubble type into eight classes provided by the NED: E, S0, Sa, Sb, Sbc, Sc, Scd, and Irr (irregular). The SNe are classified as Type Ia, SE, and Type II. SNe~Ia comprise all subtypes of these objects.  Stripped envelope (SE) SNe include Types Ib, Ib/c, and Ic; although they show evidence of envelope stripping, SNe~IIb are classified by Leaman et al. (2011) as Type II (Graur et al. 2017).   CC SNe are comprised of SE and Type II SNe.  Leaman et al. (2011) provide data from this survey on the position of each SN within its host galaxy, as well as parameters describing the apparent shape of the galactic disk.  Control times for the LOSS sample galaxies are adopted from prior work (Li et al. 2011b; Graur et al. 2017a). Using these two datasets, SN rates can be calculated for specific regions of a galaxy defined by galactocentric radius. We use this approach to determine the SN rate in the outskirts of spiral galaxies, which we define as the region of the galaxy beyond the optical radius ($r > R_{25}$ in the $B$ band, where the B-band surface brightness reaches mag 25 per square arcsecond).

The data of Leaman et al. (2011) contain the offset coordinates of each SN from the center of its host, as well as each galaxy's position angle, major-axis length, and minor-axis length. The major and minor axes are measured from the ellipse which fits the boundary of mag 25 per square arcsecond in the $B$ band. Using these parameters, the offset vector can be rotated and scaled to produce the vector as it would appear if the host galaxy were viewed face-on. The result is the vector

\begin{equation}
O''_x=\frac{d_{1}}{d_{2}}[\rm x''cos(PA)-y''sin(PA)],
\end{equation}

\begin{equation}
O''_y=x'' \rm sin(PA)+y''\rm cos(PA),
\end{equation}
\noindent
where $x''$ is the right ascension offset, $y''$ is the declination offset, PA is the position angle, $d_{1}$ is the major-axis length, and $d_{2}$ is the minor-axis length. Dividing the length of this adjusted vector by the semi-major axis length produces the galactocentric radius of the SN, normalized here to $R_{25}$:

\begin{equation}
\frac{R_{SN}}{R_{25}}=\frac{\sqrt{{O''}_x^2 + {O''}_y^2}}{(1/2) d_1}.
\end{equation}

\begin{figure}[h]
\begin{center}
\includegraphics[scale=0.55]{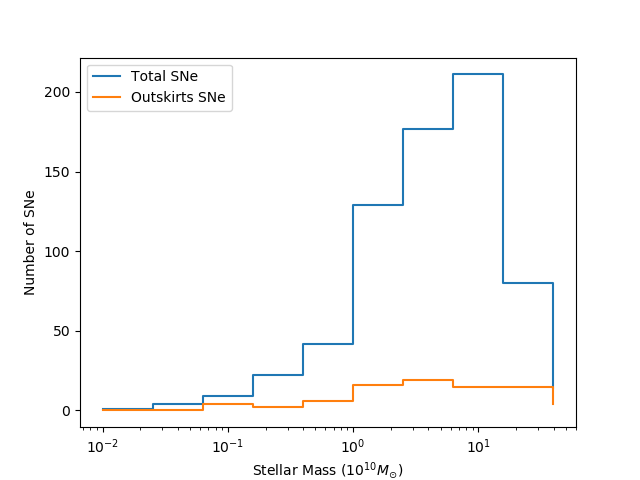} 
\caption{The number of SNe as a function of stellar mass of the galaxy, shown here for the galaxy as a whole (in blue) and in the outskirts (orange).}  \label{f:snefreq}
\end{center}
\end{figure}

Although this is a crude method of distance measurement, it is sufficient to determine which SNe explode beyond the optical radii of galaxies.  Figure \ref{f:snefreq} shows the distribution of SNe vs. the galaxy stellar mass in the sample, for the galaxy as a whole, and for the outskirts.  As is clear, the total number of SNe in the outskirts of spiral galaxies  is comparable to that in dwarf galaxies as a whole.   For dwarf galaxies, we selected galaxies with a stellar mass $< 10^{9}\,M_{\odot}$ (the mass range covers $10^{6} - 9.95 \times 10^{8}\, M_{\odot}$).  For spirals, we selected all spiral galaxies from the sample with stellar masses in the range $10^{6} -  2.8 \times 10^{12}\,M_{\odot}$.



  
 
                     







Graur et al. (2017) have provided the control times and stellar masses for each of the galaxies in the LOSS survey. When these data are combined with the parameters from the data gathered by Leaman et al. (2011), SN rates may be calculated with respect to galactocentric radius.  We calculate SN rates using the mass-normalized control-time method. The specific SN rate for a sample of galaxies, $R_{\rm sample}$, is given by 

\begin{equation}
\label{eq:sneratespec}
{\it \rm R}_{\rm sample}=\frac{N}{\Sigma t_{\rm ci} \cdot M_{*, i}},
\end{equation}
\noindent
where $N$ is the number of SNe of a certain type in the sample, $t_{\rm ci}$ is the control time of the {\it i}-th galaxy, and $M_{*,i}$ is the stellar mass of the {\it i}-th galaxy (Graur et al. 2017).  The specific SN rate increases as a galaxy's stellar mass decreases (Leaman et al. 2011). If the mass of a galaxy is known along with its specific SN rate, then the number of SNe per year can be calculated. We represent the specific SN rates in SNuM units, where 1 SNuM is $10^{-12}$\, SN\, ${\rm yr}^{-1}\,M_{\odot}^{-1}$.

The masses of the galaxies are calculated using the $B$ and $K$ magnitudes (Leaman et al. 2011; Graur et al. 2017). These magnitudes are not available for all galaxies in the sample. Leaman et al. (2011) avoid this problem by excluding these galaxies from the rate calculations, and we use the same approach to determine our SN rates. In order to increase the sample size of galaxies for the SN rate calculation, Graur et al. (2017a) extrapolated the galaxy stellar masses using one of the magnitudes. The SN rates that we have defined could be improved using a similar approach.

We use Eq. \ref{eq:sneratespec} to determine the number of SNe of a certain type that occur in the outskirts of galaxies. We calculate these rates using the same control times and stellar masses, but counting only the SNe with a galactocentric radius greater than $R_{25}$.  The control time for a given galaxy depends on the type of SN. We determine the total rate by adding the rates of each SN type:

\begin{equation}
\rm R_{\rm Total}=R_{Ia}+R_{SE}+R_{II}.
\end{equation}
Below, we calculate this rate for the outskirts of spiral galaxies, and compare it to the rate of SNe in dwarf galaxies.  The control times for each SN type in a given galaxy are comparable, so we also average the control times to get a representative control time for all SN types, which has a smaller uncertainty.  The rates determined by Graur et al. (2017) are similar to those of earlier work (Mannucci et al. 2005).  As in Leaman et al. (2011), rates are cited for the ``optimal" sample, excluding SNe that occur in early-type galaxies with small radii and in highly inclined late-type galaxies.  

\section{Results}

The radial distribution of SNe in the LOSS sample is depicted in Figure \ref{f:snerad}.  Leaman et al. (2011) attribute the small number of SNe near the center of the galaxy to extinction effects from the galactic bulge. Aside from that, the number of SNe declines as a power law as the galactocentric radius increases, but a significant number of SNe can be found beyond the optical radius; in the total sample of galaxies cross-referenced with SNe, there are 93 SNe beyond $R_{25}$, compared to an overall total of 714 SNe.

\begin{figure}[h]
\begin{center}
\includegraphics[scale=0.55]{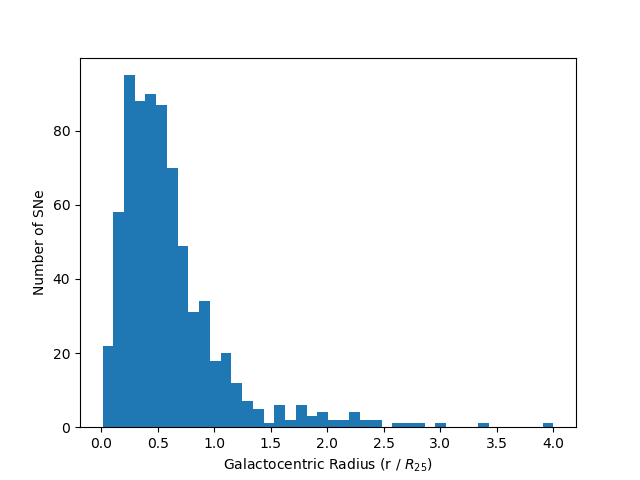}\label{f:snerad}
\caption{The number of SNe as a function of galactocentric radius (normalized here to $R_{25}$) .}  \label{f:snerad}
\end{center}
\end{figure}

We measured the dependence of the specific SN rate vs. mass (following Graur et al. 2017) in the outskirts to compare it with the SN rates overall.  Figure \ref{f:sneout} is produced by grouping galaxies by their stellar mass into bins, and both the rate and uncertainties are depicted (shown in red).   As done by Graur et al. (2017), the fixed bins are limited to the mass range $10^{8} - 10^{12}~\rm M_{\odot}$.  The sliding bin line in blue is plotted by using the same bin size as the fixed bins, but the bin slides along the abscissa and calculates the SN rate of the galaxies it contains. The gray shaded region shows the 68\% Poisson uncertainties for this calculation. We fit a power law to the fixed bins, which is plotted in green.  As has been found earlier (Graur et al. 2017), the specific SN rate declines with stellar mass. 

\begin{figure}[h]
\begin{center}
\includegraphics[scale=0.55]{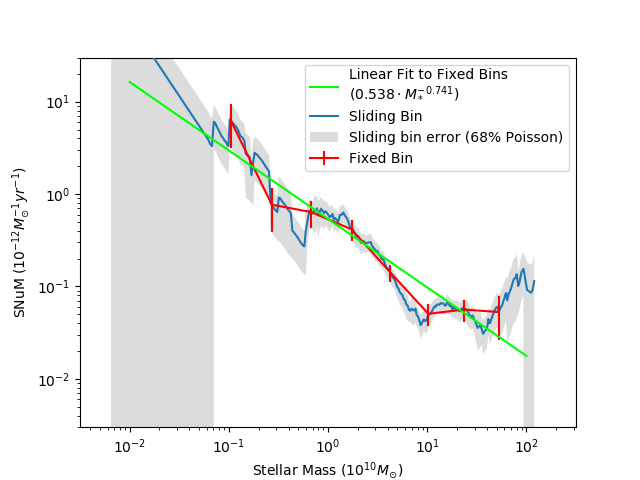}\label{f:sneout}
\caption{The specific SN rate in the outskirts of galaxies as a function of stellar mass, overplotted with a fitting function.}  \label{f:sneout}
\end{center}
\end{figure}



  


Fitting a power law to the fixed bins in Figure \ref{f:sneout} produces a specific SN rate in the outskirts given a galaxy's stellar mass,

\begin{equation}
{\rm R_{SN,outskirts}} = (1.38 \times 10^7)\, \left(\frac{M_{\star}}{M_{\odot}}\right)^{-0.74}\, {\rm SNuM}. \\ 
\end{equation}
Multiplying this specific rate by the galaxy's stellar mass produces the number of SNe per year,
\begin{equation}
{\rm SNe}~ {\rm yr}^{-1} = (1.38 \times 10^{-5})\, \left(\frac{M_{\star}}{M_{\odot}}\right)^{0.26}.  \\
\end{equation}



\begin{table*}
\centering
        \caption{SN Counts For Different SN Types$^a$}

          \begin{tabular}{@{}|l| |l | c | c | c | c || @{}}
          \hline
           &        All Types 	                              &  	Type Ia SNe         &   SE SNe               &   Type II SNe                       &   CC SNe/Type Ia SNe \\	
\hline

Total Galaxy Sample             &           714               &  $274 \pm 16$       & $116 \pm 11$        &  $324 \pm 18$                     &  $1.6 \pm 0.1$ \\

Galaxy Outskirts                 &           93                & $48 \pm 7$            & $10 \pm 3$            &   $35 \pm 6$                         & $0.9 \pm 0.2$ \\\

Spiral Outskirts                   &          73                   & $35 \pm 6$           & $10 \pm 3$             & $28 \pm 5$                           & $1 \pm 0.3$  \\ 

Spiral Inner                        &         563                 & $196 \pm 14$        & $96 \pm 10$         & $271 \pm 16$                       & $1.9 \pm 0.2$ \\

Dwarf galaxies                   &         11                    & $4 \pm 2$             & $3 \pm 2$               & $4 \pm 2$                             & $1.8 \pm 1.3$ \\

\hline

\end{tabular} 

$^a$The first column gives the SN counts for all types of SNe, the second column only for SNe~Ia, the third column only for SE~SNe, and the fourth column only for SNe~II. The fifth column gives the ratio of core-collapse (CC) SNe (SE and Type II) to SNe~Ia.

\end{table*} 

\vspace{0.3in}

The size of the spiral galaxy sample is 8,348 galaxies, which contains 73 SNe in the outskirts. The mean mass of the sample is $6.96 \times 10^{10}\,M_{\odot}$. The specific SN rate for all types of SNe is $0.035 \pm 0.007$ SNuM. The error in this rate is the 68\% Poisson uncertainty. Multiplying this specific rate by the mean mass of this sample produces a SN rate of $2.5 \pm 0.5$ SNe per millennium in spiral outskirts.

We also calculated the rate for dwarf galaxies. We collected a sample of dwarf galaxies by selecting the galaxies with a stellar mass less than $10^9\,M_{\odot}$.  This selection consists of 432 dwarf galaxies, which contain 11 SNe. The mean mass of the dwarf galaxy sample is $4.48 \times 10^{8}\,M_{\odot}$. The specific SN rate for this sample is $8.9 \pm 4.9$ SNuM. Multiplying this specific rate by the mean mass of this sample produces a SN rate of $4.0 \pm 2.2$ SNe per millenium.  For this "optimal" sample, all the dwarf galaxies are dwarf spirals. 

Averaging the control times over different SN types in a given galaxy lowers the margin of error but does not change the rates appreciably.  In this case, we obtain $2.6 \pm 0.3$ SNe per millenium and $4.0 \pm 1.1$ SNe per millenium for the outskirts of spirals and dwarf galaxies, respectively.   The rate in the outskirts of galaxies can be compared to the rate in spiral galaxies as a whole, which is $22.1 \pm 0.8$ SNe per millenium, i.e., the rate in the outskirts is about one-tenth that of the galaxy as a whole.  The rate of CC SNe in dwarfs is $2.6 \pm 1.5$ SNe per millenium, and in spiral outskirts the rate is $1.5 \pm 0.15$ SNe per millennium.   This suggests that a typical dwarf galaxy and a typical spiral outer disk are equally likely to produce the kind of dying massive stars that ultimately collapse to produce the massive merging black holes detected by LIGO/Virgo.

Finally, the relative distribution of different types of SNe in dwarf galaxies and spiral outskirts should inform our understanding of how these different environments likely contribute to the observed LIGO/Virgo detection rates.  Table 1 summarizes the numbers of the low-mass SNe~Ia and CC SNe (SE~SNe and SNe~II), as well as the relative ratio of CC SNe to SNe~Ia in the total galaxy sample, in galactic outskirts, in the outskirts of spirals, in the inner regions of spirals, and in dwarf galaxies.  The relative ratio of CC SNe to SNe~Ia is of order unity (to within the uncertainties; here we propagate the errors by adding them in quadrature) for the outskirts of spiral galaxies and dwarf galaxies.  This ratio in the inner regions of spiral galaxies is almost a factor of two higher.  

\section{Discussion and Conclusions}

The number of SNe (for the optimal sample) formed in spiral outskirts is 73 SNe out of a sample of 8348 spiral galaxies, while a total of 11 SNe are formed in the sample of 432 dwarf galaxies.  Leaman et al. (2011) note that irregular galaxies, which are correlated with low-mass galaxies (Kelvin et al. 2014), are underrepresented in the galaxy sample. However, given that there is relatively little stellar mass in dwarf galaxies, it is unlikely that the LOSS survey would have missed a significant fraction of SNe (Leaman et al. 2011).   The measurements of the SN rate in dwarf galaxies (and in outer disks) could be improved using a volume-limited sample, which would provide a more robust determination of the SN rate in low-luminosity galaxies.  

We estimate that $\sim 10$ SNe in spiral-galaxy outskirts were missed because of KAIT's small FoV; this corresponds to a 8.4\% error, smaller than the statistical error of 11.4\%. We assumed the radial distribution of SNe out to $4 \times R_{25}$ in galaxies with major axes smaller than KAIT's FoV is representative of the SN population as a whole, and used that to extrapolate results from galaxies with larger major axes.  

Owing to KAIT's small field of view (FoV; $6.7' \times 6.7'$) and the magnitude-limited ($\sim 19$\,mag) nature of the LOSS survey, the sample of dwarf galaxies is not complete (Leaman et al. 2011).   Due to the lack of completeness of dwarf galaxies in the LOSS survey, it is difficult to compare the contribution of the total number of CC~SNe formed in dwarf galaxies in some representative volume to that formed in spiral-galaxy outskirts, but we give a preliminary estimate here.  Following Li et al. (2011), SN rates are normalized to the $K$-band luminosities, and converted into volumetric rates by multiplying them by luminosity densities of the galaxy type (where galaxy type is specified by luminosity range), from Kochanek et al. (2001).   Restricting the stellar mass in dwarfs to $< 10^{9}\,M_{\odot}$ (i.e., a luminosity cut that corresponds to this mass cut) produces a volumetric rate that is $31,000 \pm 18,000$\,SNe\,Gpc$^{-3}$\,yr$^{-1}$, while the volumetric CC~SN rate for spiral outskirts is $7600 \pm 1700$\,SNe\,~Gpc$^{-3}$\,yr$^{-1}$.   For spirals, we selected all spiral galaxies in our optimal sample.  We obtain the luminosity cut for dwarf galaxies by multiplying our mass cut of $10^{9} M_{\odot}$ for dwarf galaxies by the light-to-mass ratio that is obtained when we use the B, K magnitudes to determine stellar masses (as described in \S 2).  One caveat is that the luminosity cut that we obtain from this mass cut by multiplying by the light-to-mass ratio, may be affected by the variation in the stellar light-to-mass ratio, which can be uncertain by a factor of few when using stellar population synthesis models (Portinari et al. 2004) or even higher (Bershady et al. 2010).  Here, we have followed prior work on the LOSS survey by obtaining stellar masses from the B and K magnitudes only.

A puzzling observation of outer H~I disks is that the velocity dispersion close to $R_{25}$ is about 10\,km\,s$^{-1}$, which Tamburro et al. (2009) note would require an unrealistic amount of SN feedback efficiency if SNe were entirely responsible for maintaining this velocity dispersion.  They calculated the energy input rate using an estimate of the SN rate in the outskirts, assuming an initial mass function to determine the number of stars in the outskirts that terminate as CC SNe.  We find that the ratio of the SN rate in the outskirts is about one-tenth that of the rate in the spiral galaxy as a whole, comparable to the ratio of the star-formation rates in the outer and inner galaxy (Bigiel et al. 2010).  This suggests that the SN rate estimated from the star-formation rate (Tamburro et al. 2009) should be approximately correct, and that an additional energy source such as the magnetic rotational instability may be needed (Sellwood \& Balbus 1999).


The broad agreement between the SN rates per galaxy in the outskirts of spiral galaxies and in dwarf galaxies suggests that the spiral outskirts are a region that deserves closer research.  The ratio of core-collapse SNe to SNe~Ia is comparable in both spiral outskirts and in the inner regions (but is higher in the inner regions by a factor of $\sim$ 2), as well as in dwarf galaxies, with a large uncertainty for dwarfs.  We find that dwarf galaxies contribute $31,000 \pm 18,000$\,SNe\,Gpc$^{-1}$\,yr$^{-1}$ to the volumetric rate, while spiral outskirts produce $7600 \pm 1700$\,SNe\,~Gpc$^{-1}$\,yr$^{-1}$.  Given the small numbers of SNe in dwarf galaxies in the LOSS sample, our conclusions about dwarf galaxies are approximate.  Deeper surveys would give us a more complete picture of the star-formation rate and stellar mass in the outskirts and in dwarf galaxies, which would further inform our understanding of the contribution of galactic outskirts and dwarf galaxies to the LIGO/Virgo detection rates. 

\bigskip
\bigskip

S.C. gratefully acknowledges support from the National Science Foundation (NSF) under grant AST-1517488, and from NASA grant NN17AK90G.   O.G. is supported by an NSF Astronomy and Astrophysics Fellowship under award AST-1602595. A.V.F. is grateful for financial assistance from NSF grant AST-1211916, the Christopher R. Redlich Fund, the TABASGO Foundation, and the Miller Institute for Basic Research in Science (U.C. Berkeley).  His work was conducted in part at the Aspen Center for Physics, which is supported by NSF grant PHY-1607611; he thanks the Center for its hospitality during the supermassive black holes workshop in June and July 2018.

KAIT and its ongoing operation were made possible by donations from Sun Microsystems, Inc., the Hewlett-Packard Company, AutoScope Corporation, Lick Observatory, the NSF, the University of California, the Sylvia \& Jim Katzman Foundation, and the TABASGO Foundation. Research at Lick Observatory is partially supported by a generous gift from Google.



\end{document}